\documentclass{vgtc}                          % final (conference style)
\graphicspath{{figures/}{pictures/}{images/}{./}} % where to search for the images

\usepackage{times}                     % we use Times as the main font
\usepackage{tabu}                      % only used for the table example
\usepackage{booktabs}                  % only used for the table example
\usepackage{lipsum}                    % used to generate placeholder text
\usepackage{mwe}                       % used to generate placeholder figures

\usepackage{mathptmx}                  % use matching math font

\onlineid{0}

\vgtccategory{Research}

\vgtcinsertpkg

\title{User-Guided Force-Directed Graph Layout}

\author{Hasan Balci\thanks{e-mail: hasan.balci@nih.gov}\\ %
        \scriptsize National Library of Medicine, Bethesda, MD, USA %
\and Augustin Luna\thanks{e-mail: augustin.luna@nih.gov}\\ %
     \scriptsize National Library of Medicine, Bethesda, MD, USA %
}

\teaser{
  \centering
  \includegraphics[width=\linewidth]{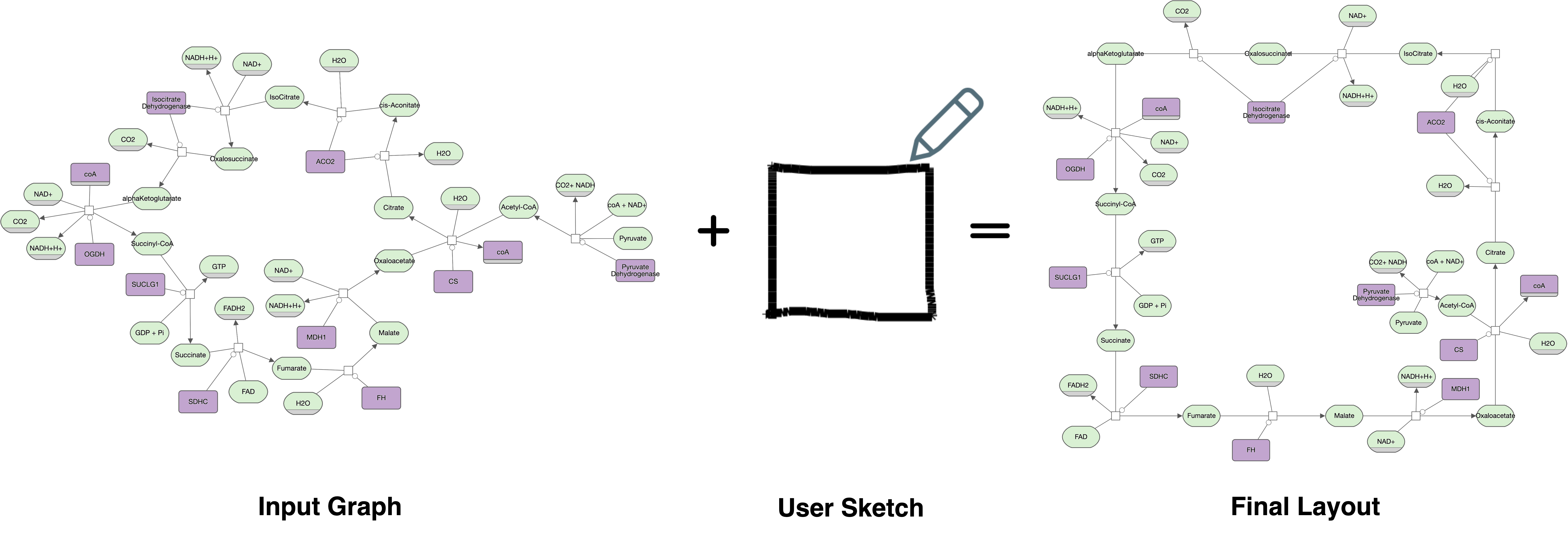}
  \caption{Starting from an input graph (left) and a user-provided sketch (middle), our approach produces a final layout (right) that aligns with the intended shape while maintaining the graph’s connectivity and semantics.}
  \label{fig:teaser}
}

\abstract{
Visual analysis of relational data is essential for many real-world analytics tasks, with layout quality being key to interpretability. However, existing layout algorithms often require users to navigate complex parameters to express their intent. We present a user-guided force-directed layout approach that enables intuitive control through freehand sketching. Our method uses classical image analysis techniques to extract structural information from sketches, which is then used to generate positional constraints that guide the layout process. We evaluate the approach on various real and synthetic graphs ranging from small to medium scale, demonstrating its ability to produce layouts aligned with user expectations. An implementation of our method along with documentation and a demo page is freely available on GitHub at https://github.com/sciluna/uggly. 
} % end of abstract


\begin{document}


%%\firstsection{Introduction}

\maketitle

\section{Introduction} %for journal use above \firstsection{..} instead
Graphs are a widely used approach to visualize relational data, representing objects and the links between them. An important aspect of this visualization is the layout, which plays a crucial role in understanding complex structures and relationships. While a well-organized arrangement of nodes and edges improves readability and analysis, a poor layout causes users to spend up to 25 percent of their time on manual layout adjustments~\cite{klauske_effizientes_2012}. To address this, numerous graph layout algorithms, such as force-directed, spectral, and hierarchical, have been developed so far, each designed to increase readability based on user needs and graph structure~\cite{di_battista_graph_1999,tamassia_handbook_2013}. However, these algorithms often lack flexibility in accommodating user preferences. Although they offer some customization options, users typically have limited direct control over the final layout beyond parameter tuning. Hence, guiding layouts intuitively without extensive manipulation of node positions manually remains a key challenge, especially for non-expert users.

This paper introduces a user-guided force-directed layout approach that lets users specify their desired layout structure by drawing sketches (e.g., rectangle, L-shape). Our method leverages a well-established image analysis technique, skeletonization (i.e., medial axis transform), to interpret the user sketch and uses the extracted structural information to guide force-directed layout algorithms with placement constraint support. It works well for small to medium-sized graphs and generates visually effective layouts aligned with user intent.

The contributions of this work are two-fold:
\begin{itemize}
    \item A novel interactive approach to force-directed layout generation that enhances user control over the graph layout without requiring expertise.
    \item An exploration of how standard image analysis methods can help close the gap between user intent and automatic layout algorithms, an area that has not been widely studied.
\end{itemize}

\section{Related Work}
Force-directed layout algorithms~\cite{tamassia_handbook_2013} are widely adopted due to their aesthetic results and ease of implementation. To better accommodate user needs, variants such as CoLa~\cite{dwyer_ipsep-cola_2006,dwyer_topology_2009,dwyer_scalable_2009} and fCoSE~\cite{balci_fcose_2022} extend these methods with placement constraints, such as aligning node sets vertically or horizontally or positioning specific nodes relative to each other. However, while these options work well for local adjustments, modifying the global structure of the graph often requires defining a large number of constraints, which is a time-consuming task for users. Efforts to simplify constraint specification include setCoLa~\cite{hoffswell_setcola_2018}, which introduces a domain-specific language for specifying high-level constraints, and the study by Petzold et al.~\cite{petzold_interactive_2023} which proposes an interactive system for defining constraints dynamically. Although these methods conceptually align with our work, they still demand technical expertise (e.g., learning a language or manually defining constraints), limiting their accessibility, particularly to non-expert users.

Beyond constraint-based options, only a few studies have explored interactive techniques that allow user guidance in graph layouts. Some studies~\cite{mi_interactive_2016,trethewey_interactive_2009} have worked on efficiently recalculating layouts after users manually drag nodes. Sketch or touch-based interactive techniques enable swift and direct control over graph elements, allowing free manipulation of nodes in graph data analysis~\cite{schmidt_set_2010,gladisch_mapping_2015}. Closer to our approach, Guchev and Gena~\cite{guchev_sketch-based_2017} investigate refining a force-directed graph layout using a pen-centric sketching approach, where an artificial neural network with backpropagation is used to interpret predefined gestures to adjust the layout. The drawbacks of this method are that it can only recognize a predefined set of gestures, allowing only circular and linear adjustments of selected nodes, and necessitates a training process for gesture recognition. These constraints highlight the need for more flexible and adaptive approaches to interactive graph layout design.

Skeletonization is a well-established image analysis technique~\cite{saha_survey_2016} that reduces binary images to their medial axis, producing a thin, connected structure that preserves the original shape and topology. It is widely used in applications such as shape matching~\cite{sundar_skeleton_2003}, shape classification [17], and character recognition~\cite{yu_skeleton-based_2008,sarnacki_character_2018} where preserving structural features is essential. In this work, we apply skeletonization to interpret user-drawn sketches and extract structural information in the form of line segments to guide graph layout, a novel use case that, to the best of our knowledge, has been largely unexplored.

Our approach proposes a new perspective to tackle the limitations of current methods by leveraging skeletonization to extract line segments from user-drawn sketches, aiding in the generation of placement constraints. This eliminates the need for technical expertise, as users guide the layout intuitively via sketches rather than manual constraint definitions or specialized languages. We also support layout creation from scratch with flexible sketch inputs. By automating the interpretation of user sketches through classical image analysis, our method provides an accessible and scalable solution for generating layouts of small to medium-sized graphs.

\begin{figure}[t]% specify a combination of t, b, p, or h for top, bottom, on its own page, or here
  \centering % avoid the use of \begin{center}...\end{center} and use \centering instead (more compact)
  \includegraphics[width=0.45\textwidth, alt={A diagram showing algorithm overview.}]{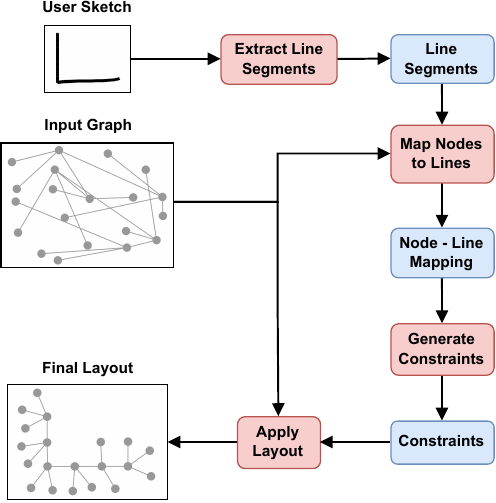}
  \caption{%
      Algorithm overview. Given an input graph and a user-drawn sketch, our approach first applies skeletonization followed by polyline simplification to extract line segments. Graph nodes are then mapped to these segments, forming a node-line mapping used to generate placement constraints. These constraints guide the final layout generation.
  }
  \label{fig:methodology}
\end{figure}

\section{Methodology}
Given an input graph and a user-drawn sketch, our approach follows four steps to generate a final layout that aligns with the user’s intent (\cref{fig:methodology}). First, line segments are extracted from the sketch using skeletonization followed by polyline simplification. In the second step, graph nodes are assigned to these segments, resulting in a node-line mapping.  Based on this mapping, we generate layout-specific placement constraints in the third step. Lastly, a force-directed layout algorithm is applied using the original graph and the generated constraints to obtain the final layout.

\subsection{Line Segment Extraction}
\begin{figure}[h]
  \centering
  \begin{tabular}{@{}c@{}}
  	\centering
  	\includegraphics[width=0.32\columnwidth]{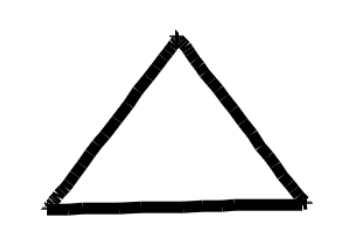}
  	\label{fig:skeleton_a}
  \end{tabular}%
  \hfill%
  \begin{tabular}{@{}c@{}}
  	\centering
  	\includegraphics[width=0.32\columnwidth]{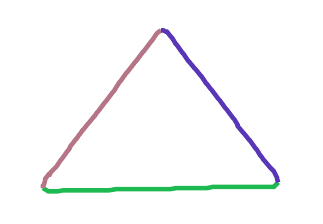}
  	\label{fig:skeleton_b}
  \end{tabular}%
  \hfill%
  \begin{tabular}{@{}c@{}}
  	\centering
  	\includegraphics[width=0.32\columnwidth]{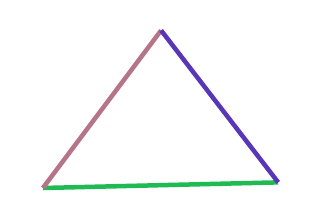}
  	\label{fig:skeleton_c}
  \end{tabular}%
  \caption{(left) User sketch (middle) Three polylines, each shown with a different color, extracted via skeletonization (right) Corresponding line segments after polyline simplification.}
  \label{fig:skeleton}
\end{figure}

We begin extracting line segments from the user’s sketch (\cref{fig:skeleton}), provided as a raster image, by first converting the image into binary format if it is not already. We then apply Huang's skeleton extraction algorithm\footnote{https://github.com/LingDong-/skeleton-tracing}, which retrieves the topological skeleton as a set of polylines. The algorithm starts by skeletonizing the binary image using the Zhang-Suen thinning method~\cite{zhang_fast_1984}, then recursively splits the image along optimal rows or columns. It extracts polylines from smaller regions by tracing borders and connecting key points to the center, ultimately merging the results into a final set of polylines (\cref{fig:skeleton}). 

These polylines often contain numerous short segments defined by their endpoints. To improve interpretability and better capture the overall structure of the sketch, we simplify each polyline using the polyline simplification method by Agafonkin\footnote{https://github.com/mourner/simplify-js} which uses a combination of Ramer-Douglas-Peucker~\cite{ramer_iterative_1972, douglas_algorithms_1973} and radial distance algorithms. This produces a more compact representation with fewer and longer polylines (\cref{fig:skeleton}).

Finally, we merge and order the line segments from simplified polylines such that the start point of each segment matches the end point of the previous one, except for the first segment. If the segments form a closed loop, the start point of the first segment coincides with the end point of the last. We represent the final output as an ordered list of line segments \( L = \{ l_i = (p_{i}, p_{i+1}) \mid i = 1, \dots, n \} \), where each point \( p_{i} = (x_i, y_i) \in R^{2}\), and consecutive segments share endpoints by construction. If the segments form a closed loop, then \(p_{n+1} = p_1\).

\subsection{Node-Line Mapping}
In this step, we map graph nodes to the extracted line segments using a heuristic approach to achieve an overall distribution that aligns with the structure of the user’s sketch. Let the input graph be \( G = (V, E) \), where \( V \) is the node set and \( E \) is the edge set. We define a subgraph \( G'=(V', E') \) where \( V^{'} = \{v \in V \mid deg(v)>1\} \) and \( E^{'}=\{(u,v) \in E \mid u \in V', v \in V'\} \); that is, we exclude nodes with degree one together with their adjacent edges as they contribute little to global structure. Note the input graph, whether directed or undirected, is treated as undirected for this step.

If the extracted line segments form a closed loop \( (p_{n+1} = p_1) \), we attempt to identify a representative cycle \( C \subseteq {V'} \) in the graph that is sufficiently long to reflect its structure. Since finding the longest cycle in a graph is an NP-hard problem~\cite{karp_reducibility_nodate}, we use an approximation strategy. Specifically, we perform a depth-first search (DFS) from each node and track the longest cycle encountered. To reduce computational overhead, we skip DFS from nodes that have already been visited, which significantly reduces redundant exploration, especially in dense graphs.

To ensure the extracted cycle is meaningful (e.g., not too small to reflect the graph’s overall structure), we accept the cycle \( C = \{v_1, v_2, \dots, v_m\} \) only if its length \( |C| \ge \tau \), where we use a threshold \( \tau = 2\sqrt{|V'|} \) based on empirical observations, though this parameter can be adjusted for flexibility. Once such a cycle is found, we distribute its nodes across the line segments in \( L \) proportionally to the length of each segment. Let \( d_i = \| p_{i+1} - p_i \| \) be the length of segment \( l_i \), and \( D = \sum_{i=1}^{n} d_i \) be the total length of the line segments. Then the number of nodes assigned to segment \( l_i \) is given by

\[
k_i = \biggl \lfloor \frac{d_i}{D} \cdot |C| \biggr \rfloor
\]

Nodes are assigned to segments in traversal order, maintaining their relative positions.

If the extracted line segments do not form a closed loop, or if no sufficiently long cycle is found, we switch to a more general mapping strategy. We perform a two-pass breadth-first search (BFS): starting from a randomly selected node \( v \in V' \), we find the furthest reachable node \( v_{far} \). A second BFS is then performed starting from \( v_{far} \), resulting in an ordered list of visited nodes \( S = \{ u_1, u_2, \dots, u_k \} \), where \( k = |V'| \). Assuming the graph has a relatively uniform degree distribution, the same proportional assignment strategy is used to distribute \( S \) over the line segments \( L \), ensuring spatial consistency with the user sketch.

We define the node-line mapping as a set of associations between each line segment \( l_i \in L \) and the ordered list of graph nodes \( N_i \in V' \) mapped to it:

\[
M = \{ (l_i, N_i) \mid i = 1, \dots, n \}
\]

where \( N_i = \{v_{1}^{i}, v_{2}^{i}, \dots \} \) represents the nodes assigned to line \( l_i \), in traversal order. Additionally, we maintain a parent map \( P : V' \to V' \cup \{ \emptyset \} \), where \( P(v) \) gives the predecessor of node \( v \) in the traversal. This mapping is later used to generate relative placement constraints.

\subsection{Constraint Generation}

In this step, we use the node-line mapping \textit{M} together with the parent map \textit{P} from the previous step to generate the placement constraints required for the force-directed layout algorithm. We implement constraint generation for the fCoSE algorithm, which supports the following constraints:

\begin{itemize}
  \item Relative placement constraints: Constraining a node's position relative to another node in either vertical or horizontal direction (e.g., \texttt{\{left: n0, right: n1\}}, \texttt{\{top: n2, bottom: n3\}}).
  \item Alignment constraints: Enforcing the alignment of two or more nodes either vertically or horizontally (e.g., \texttt{\{horizontal: [[n0, n1, n2], [n3, n4]], vertical: [[n0, n5], [n3, n6, n7]]\}}).
\end{itemize}

These constraints are sufficient to enforce the structure dictated by the user-drawn shape and can easily be adapted for use in algorithms supporting similar constraints, such as the CoLa layout.

Constraint generation is performed by processing each line segment sequentially. For each line segment \(l_{i} \in L \) from \( (x_{i}, y_{i} ) \) to \( (x_{i+1}, y_{i+1}) \), we start by determining its direction. We first compute \(\Delta x = x_{i+1} - x_{i} \) and \(\Delta y = y_{i+1} - y_{i} \). If \( |\Delta y / \Delta x| < \epsilon \), the line is classified as horizontal: \textit{l-r} if \( \Delta x > 0 \), \textit{r-l} if \( \Delta x < 0 \). If  \( |\Delta x / \Delta y| < \epsilon \), it is considered vertical: \textit{t-b} if \( \Delta y > 0 \), \textit{b-t} if \( \Delta y < 0 \). Otherwise, the line is diagonal: \textit{tl-br} if both \( \Delta x > 0 \) and \( \Delta y > 0 \), \textit{br-tl} if both are negative, \textit{tr-bl} if \( \Delta x < 0 \) and \( \Delta y > 0 \), or \textit{bl-tr} if \( \Delta x > 0 \) and \( \Delta y < 0 \). Here, \( \epsilon \) denotes a slope threshold and letters \textit{l}, \textit{r}, \textit{t}, and \textit{b} denote \emph{left}, \emph{right}, \emph{top}, and \emph{bottom}, respectively. We define the slope threshold to capture the horizontal and vertical line segments more efficiently by ignoring slight changes in the direction since the drawings are made by hand and each line segment may not be drawn linearly. We use \( \epsilon = 0.2 \) as the threshold based on empirical observations, but it can be adjusted according to the desired sensitivity.

We then generate the constraints in the following way:

\begin{itemize}
  \item Relative placement constraints: For each node \( v_{j}^{i} \in N_{i} \)
mapped to \( l_{i} \), we look for its predecessor \( q = P(v_{j}^{i}) \). Then for \( v_{j}^{i} \) and 
 \( q \) pair, we determine the constraints based on the segment’s direction.
  If \( l_i \) is horizontal or vertical (\textit{l-r}, \textit{r-l}, \textit{t-b} or \textit{b-t}), we define a single constraint accordingly. For example, if direction is \textit{l-r}, the corresponding constraint is \texttt{\{left: q, right: v\textsubscript{j}\textsuperscript{i}\}}. If \( l_i \) is diagonal ( \textit{tl-br}, \textit{br-tl}, \textit{tr-bl} or \textit{bl-tr}), we define two constraints in both horizontal and vertical directions to preserve relative positioning. For example, if direction is \textit{tl-br}, the corresponding constraints are \texttt{\{left: q, right: v\textsubscript{j}\textsuperscript{i}\}} and \texttt{\{top: q, bottom: v\textsubscript{j}\textsuperscript{i}\}}.

  \item Alignment constraints: If \( l_i \) is horizontal (\textit{l-r} or \textit{r-l}), we enforce horizontal alignment on the corresponding node sequence \( [v_1^i, v_2^i, \dots] \). If \( l_i \) is vertical (\textit{t-b} or \textit{b-t}), we instead enforce vertical alignment on the same sequence.

\end{itemize}

\subsection{Applying Layout}

Once placement constraints are generated, we first apply the force-directed layout algorithm on the original input graph, incorporating these constraints as layout options. While these constraints enforce the core structure to align with the user’s intent, one-degree nodes are positioned freely around their adjacent nodes based on the force-directed model. However, in cases where we include all nodes, except those with degree one, in constraints, the resulting layout can sometimes appear overly rigid. To address this and achieve a more natural look in the final layout, we additionally apply an unconstrained incremental layout with a small number of iterations to introduce slight adjustments, improving visual aesthetics while preserving structural consistency.

\subsection{Time Complexity}

Skeletonization runs in \(O(wh)\) where \(w \times h\) is the image resolution and polyline simplification takes \(O(nlogn)\) per polyline, where $n$ is the number of points. For node-line mapping, the cycle-based approach runs in \(O(\mid V\mid + \mid E \mid)\), avoiding redundant depth-first searches by skipping visited nodes. The fallback BFS-based mapping also operates in \(O(\mid V\mid + \mid E \mid)\), requiring two full graph traversals. Constraint generation is linear in the number of mapped nodes which is \(O(\mid V\mid)\). Finally, the layout is computed using the fCoSE algorithm, with a time complexity of \(O(\eta(\mid V\mid + \mid E \mid))\), where \(\eta\) is the number of iterations required for convergence. In practice, the pipeline is efficient and scales well with moderate graph sizes (up to a few thousand nodes and edges) and typical image resolutions of around \(512 \times 512\) pixels.

\begin{figure*}[t]
  \centering
  \begin{tabular}{@{}c@{}}
  	\centering
  	\includegraphics[width=0.38\textwidth, alt={}]{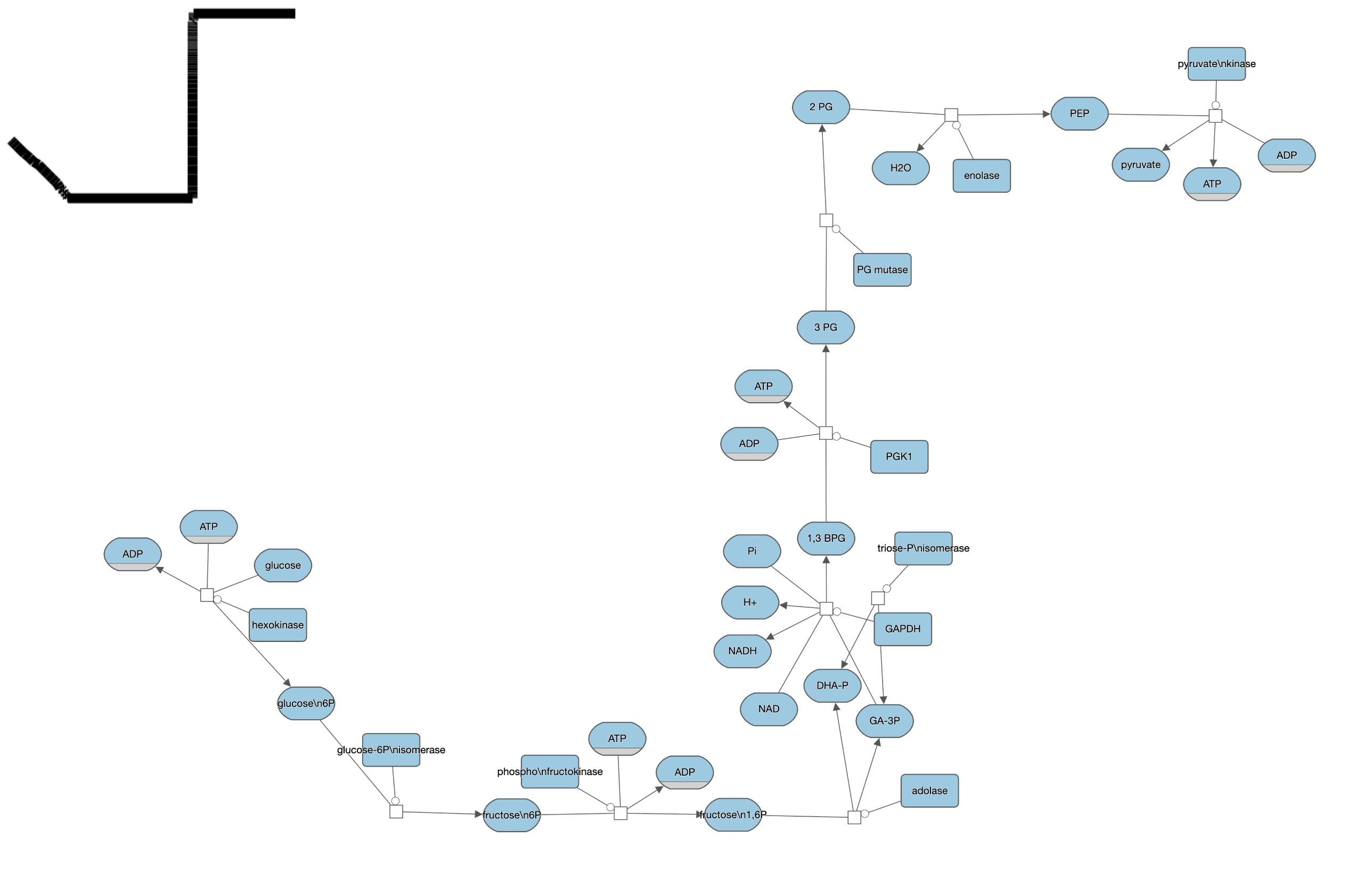} 
  	\label{fig:sample1}
  \end{tabular}%
  \hfill%
  \begin{tabular}{@{}c@{}}
  	\centering
  	\includegraphics[width=0.38\textwidth, alt={}]{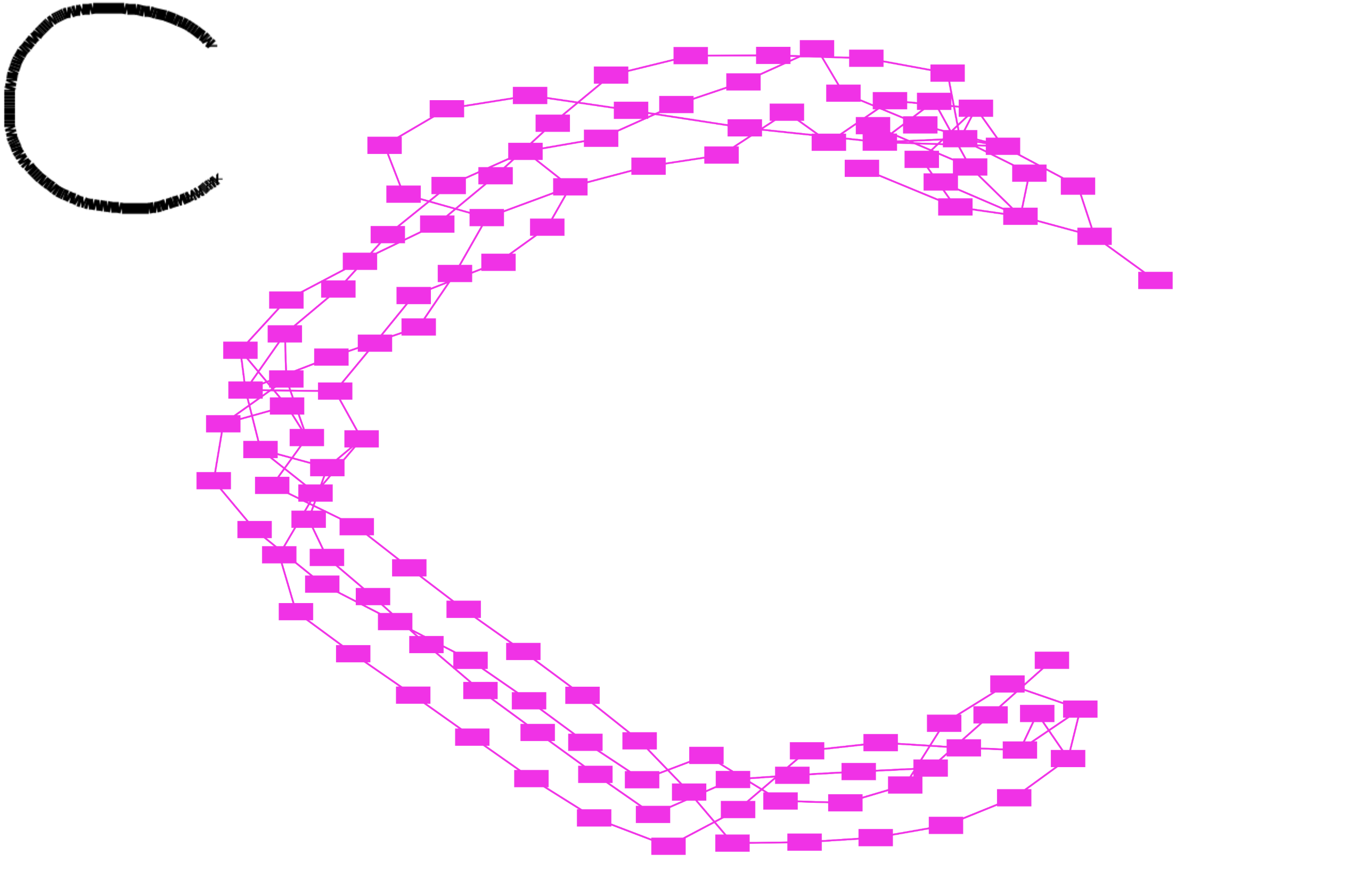}
  	\label{fig:sample2}
  \end{tabular}%
  \\%
  \begin{tabular}{@{}c@{}}
  	\centering
  	\includegraphics[width=0.38\textwidth, alt={}]{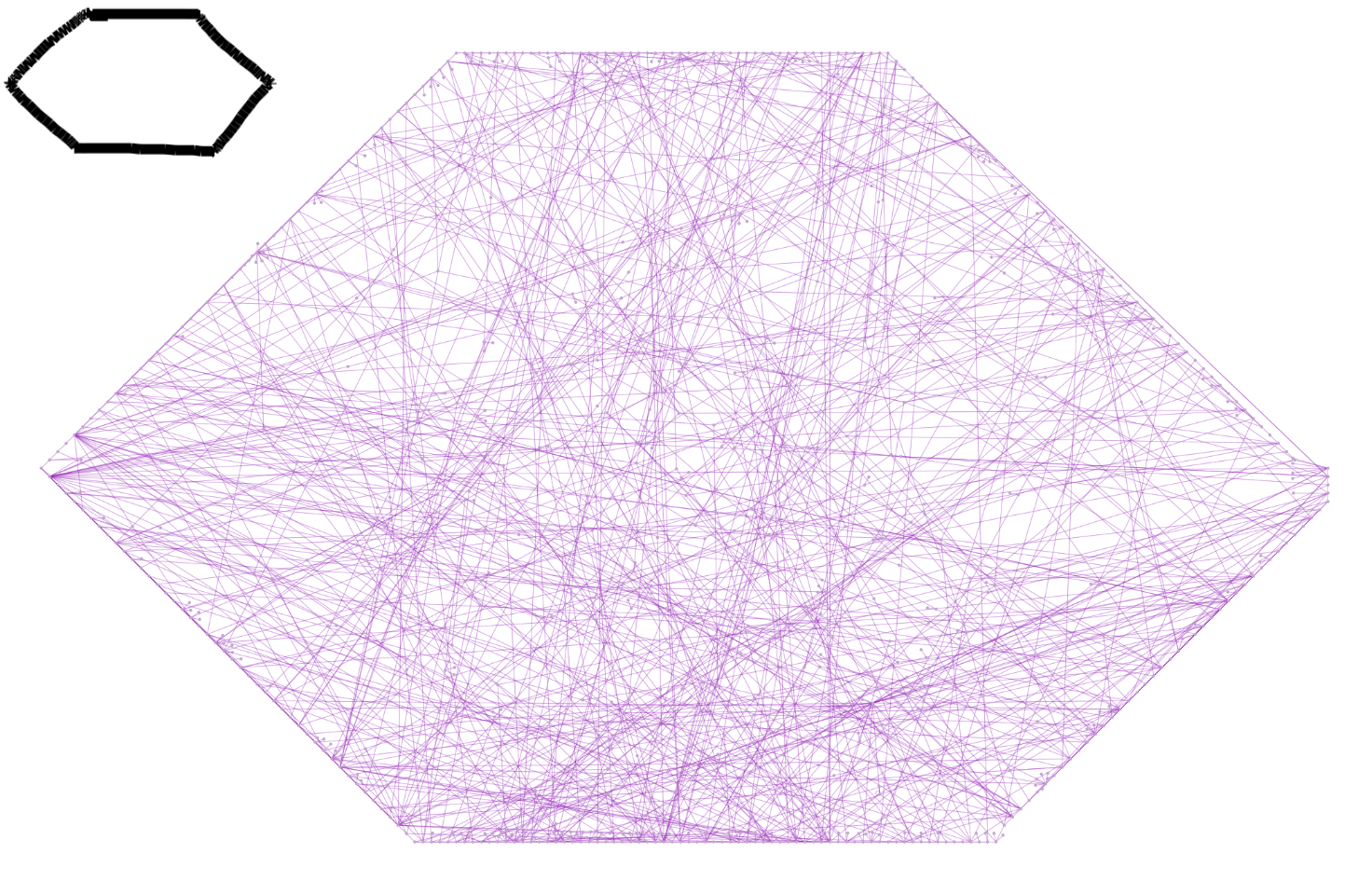}
  	\label{fig:sample3}
  \end{tabular}%
  \hfill%
    \begin{tabular}{@{}c@{}}
  	\centering
  	\includegraphics[width=0.38\textwidth, alt={}]{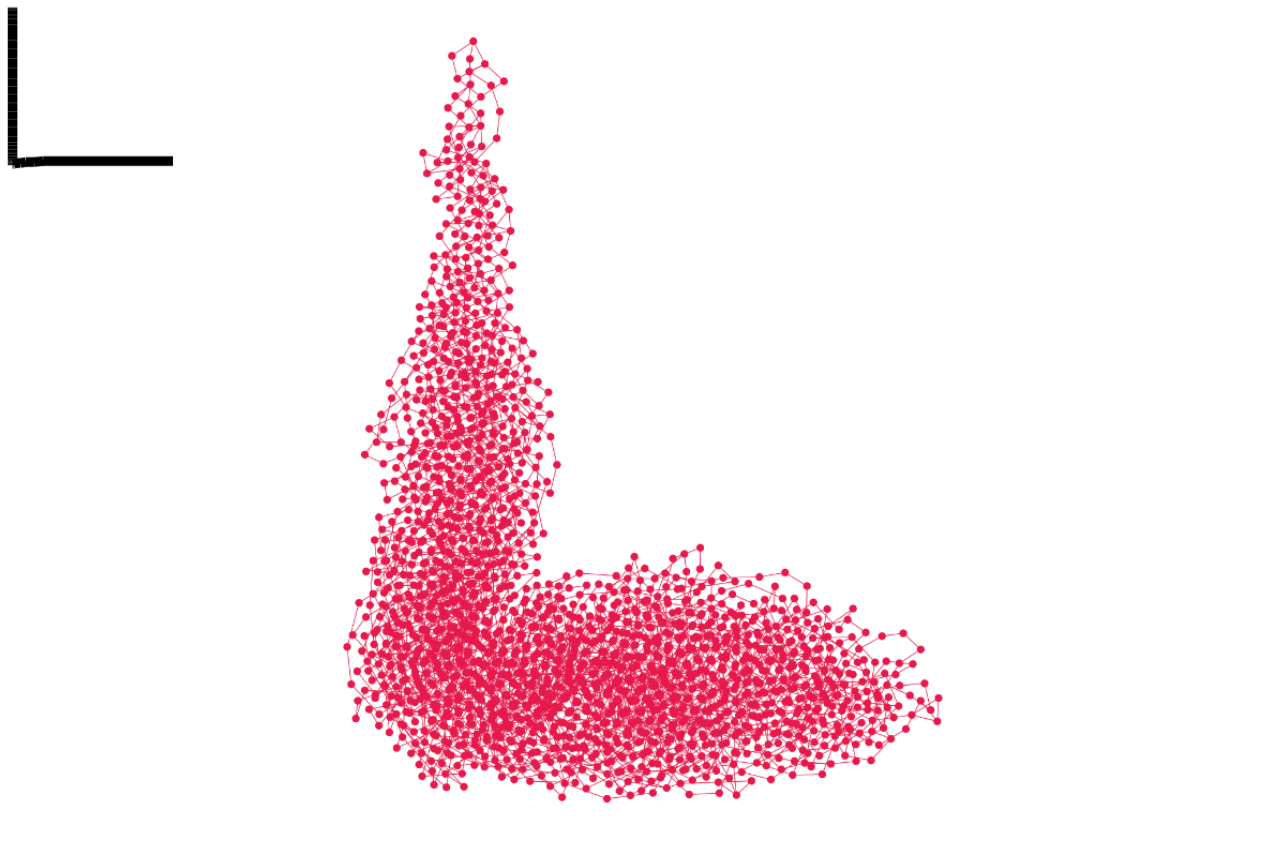}
  	\label{fig:sample4}
  \end{tabular}%
  \caption{Sample graphs each laid out with our approach by using the hand-drawn sketches shown on their upper left. (top-left) A biological network with \(|V| = 44\) and \(|E| = 44\). (top-right) A cheminformatics network from the Network Repository website~\cite{rossi_network_2015} with  \(|V| = 125\) and \(|E| = 282\) (bottom-left) A crime network from the Network Repository website~\cite{rossi_network_2015} with \(|V| = 829\) and \(|E| = 1476\). (bottom-right) A sample graph from Rome graph dataset~\cite{bridgeman_turn-regularity_2000} with \(|V| = 2000\) and \(|E| = 2512\). }
  \label{fig:samples}
\end{figure*}

\section{Results}
Our implementation is in JavaScript using the Cytoscape.js~\cite{franz_cytoscapejs_2023} graph visualization library. In the last algorithm step, we used its fCoSE layout extension as the constrained layout.

We tested our approach on a range of real-world and synthetic graphs of varying sizes and structural complexity as shown in \cref{fig:samples}. The examples include a biological network, cheminformatics and crime networks from the Network Repository~\cite{rossi_network_2015}, and a benchmark graph from the Rome graph dataset~\cite{bridgeman_turn-regularity_2000}. Across these diverse scenarios, the layouts produced closely reflect the corresponding hand-drawn sketches, demonstrating that our method scales well from small to medium-sized graphs and adapts to different user intentions. Scalability is primarily limited by the performance of force-directed layouts with constraint support.

\begin{figure}[h!]
  \centering
  \begin{tabular}{@{}c@{}}
  	\centering
  	\includegraphics[width=0.45\columnwidth, alt={}]{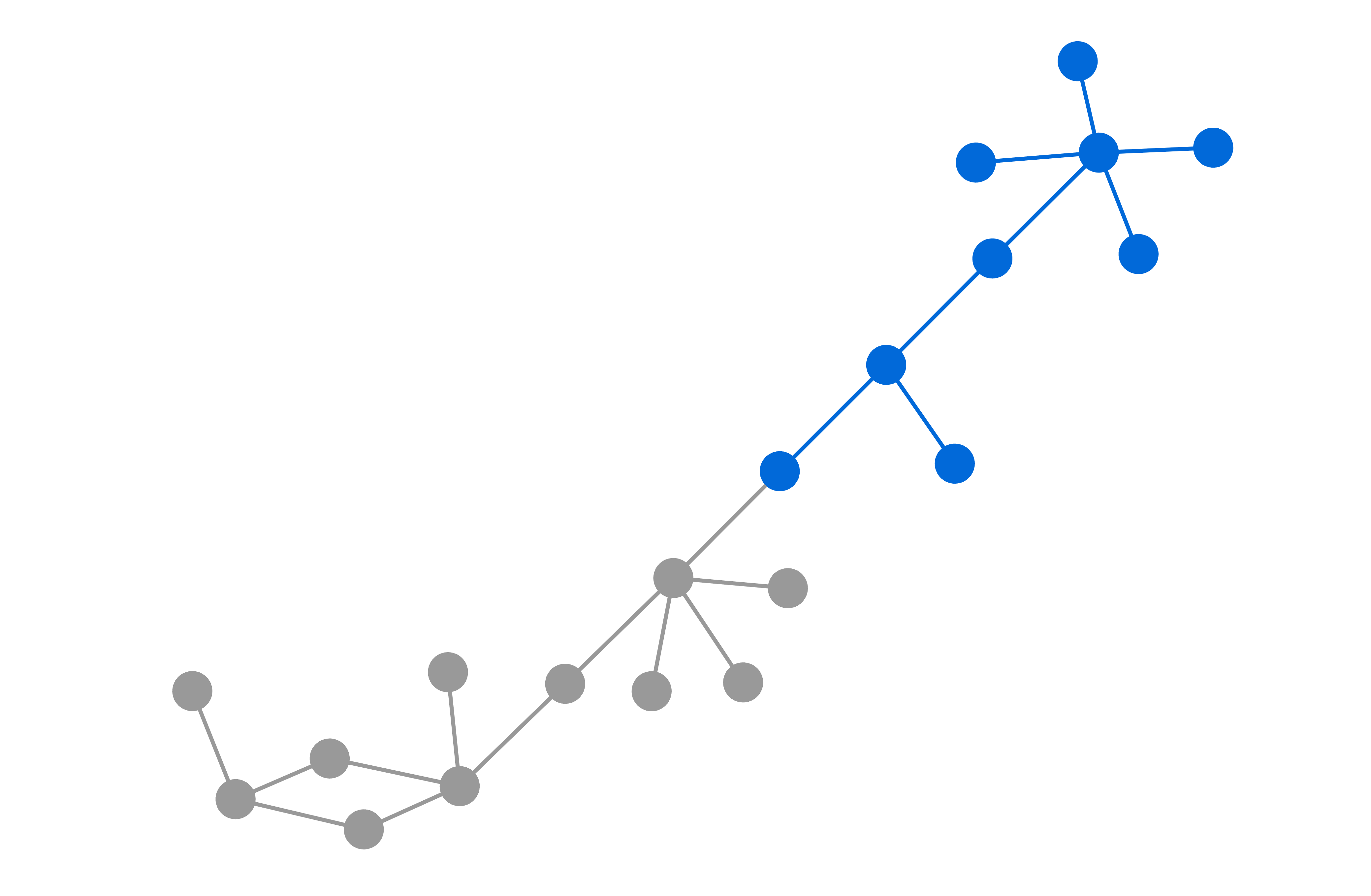}
  \end{tabular}%
  \hfill%
  \begin{tabular}{@{}c@{}}
  	\centering
  \includegraphics[width=0.45\columnwidth, alt={}]{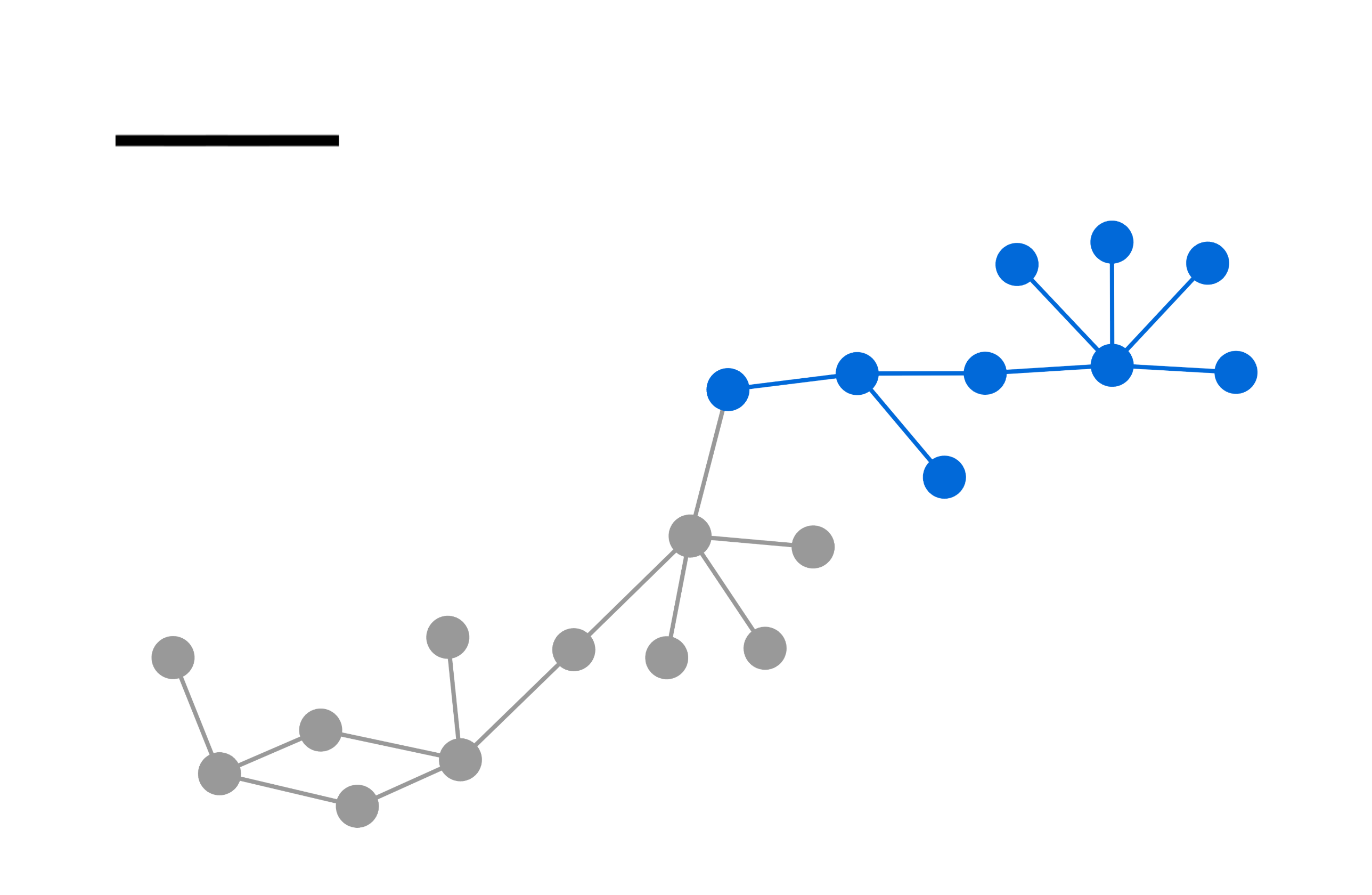}
  \end{tabular}%
  \caption{ (left) A sample graph where the selected part is indicated in blue (right) After the selected part is laid out incrementally based on a sketch indicating a horizontal adjustment.}
  \label{fig:incremental}
\end{figure}

As shown in \cref{fig:sample2} (top-right), the method tries to approximate curved paths using short line segments, which can slightly affect fidelity, though it still successfully captures the overall shape. Our method also supports incremental adjustments on selected portions of the graph in addition to the from scratch layout (\cref{fig:incremental}). 

We also explored the use of popular multimodal large language models (LLMs) for our extraction step by comparing our skeletonization/polyline simplification processing method to line extraction done solely with GPT-4o and Gemini 2.0 Flash. We generated sketch-based layouts for 160 graphs from the Rome graph dataset~\cite{bridgeman_turn-regularity_2000}, using different extraction methods. In a human evaluation on Amazon Mechanical Turk (2500 pairwise comparisons by 51 workers), the layouts generated by using our approach were preferred overall. We generated rankings of user preference using the Bradley-Terry model~\cite{bradley_rank_1952}, our approach achieved a score of 1.60, compared to 0.86 for Gemini and 0.76 for GPT-4o. Failures in the LLM-based approaches were mostly due to their instability and incorrect assumptions about the sketch content. Further details on this experiment are available on GitHub.

\section{Discussion and Future Work}
Our study presents a promising direction for intuitive, sketch-based graph layout manipulation, particularly effective for small to medium-sized graphs. Our method demonstrates that users can influence layouts with simple drawings. There are a few limitations to consider. Our method relies on the quality and continuity of user sketches—disconnected lines or loosely drawn curves can hinder accurate line extraction. While minor misalignments at segment endpoints can be addressed using a small offset threshold, highly curved or noisy sketches may still cause incorrect extraction. Currently, the method works best for sketches composed of consecutive lines and does not fully support branching or disconnected components. Expanding the system to handle more complex sketch structures, such as branching paths, would be a valuable next step. Another promising future direction is incorporating LLMs to suggest sketch templates based on the graph's topology or semantics, providing guided interactivity. LLMs could also help refine or interpret unclear or low-quality sketches before processing, improving the connection between freeform input and structural precision.

\acknowledgments{
HB and AL received funding support from the Division of Intramural Research (DIR) of the National Library of Medicine (NLM), National Institutes of Health (NIH) (ZIALM240126).}

\bibliographystyle{abbrv-doi}

\bibliography{uggly}
\end{document}